\documentclass[preprintnumbers,amsmath,amssymb,aps,prl,floatfix,lengthcheck,superscriptaddress]{revtex4}
\usepackage{graphicx}
\usepackage{braket}
\usepackage{upgreek}
\usepackage{color}

\begin{document}

\title{Precision spectroscopy of negative-ion resonances in ultralong-range Rydberg molecules}

\author{F. Engel}
\author{T. Dieterle}
\affiliation{5. Physikalisches Institut and Center for Integrated Quantum Science and Technology, Universit\"{a}t Stuttgart, Pfaffenwaldring 57, 70569 Stuttgart, Germany}
\author{F. Hummel}
\author{C. Fey}
\affiliation{Zentrum f\"{u}r optische Quantentechnologien, Fachbereich Physik, Universit\"{a}t Hamburg, 22761 Hamburg, Germany}
\author{P. Schmelcher}
\affiliation{Zentrum f\"{u}r optische Quantentechnologien, Fachbereich Physik, Universit\"{a}t Hamburg, 22761 Hamburg, Germany}
\affiliation{The Hamburg Centre for Ultrafast Imaging, Universit\"{a}t Hamburg, 22761 Hamburg, Germany}
\author{R. L\"{o}w}
\author{T. Pfau}
\author{F. Meinert}
\affiliation{5. Physikalisches Institut and Center for Integrated Quantum Science and Technology, Universit\"{a}t Stuttgart, Pfaffenwaldring 57, 70569 Stuttgart, Germany}
\date{\today}

\begin{abstract}
The level structure of negative-ions near the electron detachment limit dictates the low-energy scattering of an electron with the parent neutral atom. We demonstrate that a single ultracold atom bound inside a Rydberg orbit forming an ultralong-range Rydberg molecule provides an atomic-scale system which is highly sensitive to electron-neutral scattering and thus allows for detailed insights into the underlying near-threshold anion states. Our measurements reveal the so far unobserved fine structure of the $^3P_J$ triplet of Rb$^-$ and allow us to extract parameters of the associated $p$-wave scattering resonances which deviate from previous theoretical estimates. Moreover, we observe a novel alignment mechanism for Rydberg molecules mediated by spin-orbit coupling in the negative ion.
\end{abstract}

\maketitle

Negative ions constitute remarkable objects which have been studied intensively over the past decades \cite{Andersen2004,Pegg2004}. In contrast to neutral atoms or positively charged ions, anions are much more weakly bound by shallow and short-range potentials and typically feature only few bound states. As  a consequence, they have been proven ideal model systems for investigating the role of electron-electron correlations on their level structure \cite{Pegg2004}. More recently, the observation of excited opposite-parity bound states \cite{Bilodeau2000} has triggered renewed interest in high-resolution negative-ion spectroscopy \cite{Warring2009,Walter2014,Jordan2015} motivated by prospects to realize laser cooling for trapped anions \cite{Kellerbauer2006,Pan10}.

The fine details of the interaction potentials which determine negative-ion bound states also dictate the very low-energy quantum scattering of its neutral parent atom with a free electron \cite{Buckman1994,Bahrim2000,Bartschat2003}. Particularly, broad scattering resonances can arise when the associated negative-ion system hosts a short-lived transient state, bound by a centrifugal barrier and located just a few meV above the electron detachment limit. Accessing details of these underlying anion states such as relativistic fine-structure effects experimentally, however, is challenged by their short lifetime, low energy \cite{Lee1996}, or by selection rules in photodetachment studies starting from the negative-ion ground state \cite{Scheer1998}. In this Letter, we demonstrate a completely different route to investigate these systems by devising an ultra-sensitive microscopic scattering laboratory provided by an ultralong-range Rydberg molecule (ULRM) \cite{Greene2000,Khuskivadze2002,Bendkowsky2009,Shaffer2018}. ULRMs consist of a Rydberg atom which binds to a neutral ground-state atom inside the electron orbit via frequent low-energy scattering of the latter with the quasi-free Rydberg electron.

\begin{figure}[!ht]
\centering
	\includegraphics[width=\columnwidth]{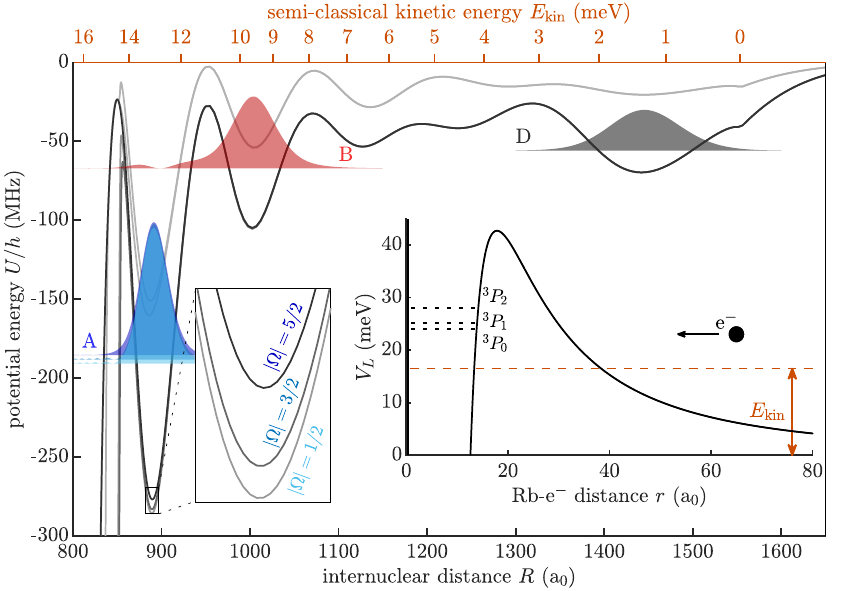}
	\caption{Anion spectroscopy in ULRMs. Molecular potential energy for the $31S$ Rydberg level owing to (triplet) Rb-e$^-$ scattering as a function of internuclear distance $R$ (black line). Vibrational wavefunctions for the molecular states A (blue), B (red), and D (black) are denoted by shaded areas. The $p$-wave dominated well ($R=890 \, a_0$) is split due to the Rb$^-$ fine-structure triplet $^3P_J$. For completeness, the gray line shows the shallow PEC originating from mixed singlet-triplet scattering (not accessed in this work). Inset: Long-range centrifugal barrier leading to the $^3P_J$ states (dotted lines) of Rb$^-$ above the electron detachment limit ($V_L=0$).}
	\label{Fig1}
\end{figure}

Here, we exploit ULRMs at an unprecedented quantitative level and demonstrate their potential to perform precise spectroscopy of negative-ion resonances at the example of the quasi-bound $^3 P_J$ state of  Rb$^-$. To this end, we identify previously unobserved molecular states, which are dominated by resonant electron-atom $p$-wave scattering, and thereby allow us to reveal the presence of relativistic spin-orbit coupling leading to a fine-structure triplet ($J \in \{0,1,2\}$). The latter was predicted theoretically for the heavy alkali metals Rb$^-$, Cs$^-$, and Fr$^-$, but so far remained experimentally inaccessible \cite{Bahrim2000,Bahrim2001}. Moreover, we observe that the presence of spin-orbit interaction aligns the ULRM even for spherically symmetric Rydberg $S$-states \cite{Hummel2019}. Our approach holds intriguing perspectives for high-resolution probing of more complex perturbers such as molecules or clusters \cite{Rittenhouse2010,Eiles2015}.

A ULRM naturally provides a highly adjustable atomic-scale system for precise studies of electron-neutral scattering at collisions energies in the meV regime \cite{Bendkowsky2009,Bendkowsky2010}. Apart from the capability to achieve very low scattering energies unfeasible to realize with free electrons, the high precision arises from the resonating Rydberg electron wave confined in the Coulomb potential, and the resulting narrow Rydberg states. Consider now the presence of a single neutral ground-state atom inside the Rydberg orbit at a distance $R$ from the Rydberg core. Quantum scattering of the electron off the perturber gives rise to a phase shift imparted on the electron wavefunction, which is detectable via a slight shift of the Rydberg electron resonance energy.

Here, we focus on ultralong-range dimers consisting of a single $^{87}$Rb ground-state atom inside the orbit of $nS_{1/2}$ Rb Rydberg states ($n=31...37$). For values of $R$ comparable to the size of the Rydberg orbit, the semi-classical electron momentum $k$ is sufficiently small so that $s$-wave scattering, as quantified by an energy-dependent (triplet) scattering length $a_s^T (k)$, dominates. This gives rise to a smoothly varying potential energy of the system as a function of $R$ which reflects the nodal structure of the Rydberg electron wavefunction (see Fig.~\ref{Fig1}) \cite{Greene2000,Khuskivadze2002,Jungen88}. For smaller values of $R$ the electron momentum increases and $p$-wave scattering can become relevant. Importantly, the $p$-wave contribution is enhanced by a shape-resonance arising from the presence of the Rb$^-$$(^3 P)$ state \cite{Hamilton2002,Niederpruem2016}, which ab-initio theoretical predictions locate about $23$ meV above the Rb - e$^-$ threshold \cite{Bahrim2000,Bahrim2001}. The resonant $p$-wave contribution leads to deep potential energy minima with decreasing $R$ when the electron kinetic energy approaches the $^3 P$ resonance. The motion of the perturber atom is dictated by the resulting potential energy curve (PEC) and quantized due to the strong radial confinement associated with the potential wells, leading to discrete vibrational dimer states. In Fig.~\ref{Fig1}, the resulting lowest lying vibrational wavefunctions are indicated for the $s$-wave ($p$-wave) dominated wells at $R=1450 \, a_0$ ($R=890 \, a_0$).

Let us now turn to the internal spin structure of the system and in particular the consequence of spin-orbit coupling in the Rb$^-$$(^3 P_J)$ state. In general, the latter gives rise to three (overlapping) anion states and consequently to a splitting of the single shape-resonance into a triplet. A theoretical treatment of this spin-orbit coupling in the context of ULRMs has been provided in Refs.~\cite{Khuskivadze2002,Eiles2017,Markson2016}. For the PEC in Fig.~\ref{Fig1}, the spin-orbit interaction leads to a splitting of the deep $p$-wave dominated potential well into three substates, while the $s$-wave dominated outer part of the PEC is essentially unaffected. The splitting is due to three different $p$-wave scattering channels associated with the $(^3 P_J)$ states, which are quantified by respective (triplet) scattering lengths $a_{p,J}^T (k)$. Each of the three split PECs is two-fold degenerate and can be associated with a different projection of the total angular momentum on the internuclear axis $|\Omega| = |m_F + m_j|$ \cite{Eiles2017}. Here, $m_j$ and $m_F$ denote magnetic quantum numbers for the Rydberg electron spin and the ground-state atom hyperfine level, respectively. We focus on the experimentally relevant PECs for $F=2$.

\begin{figure}[t]
\centering
	\includegraphics[width=\columnwidth]{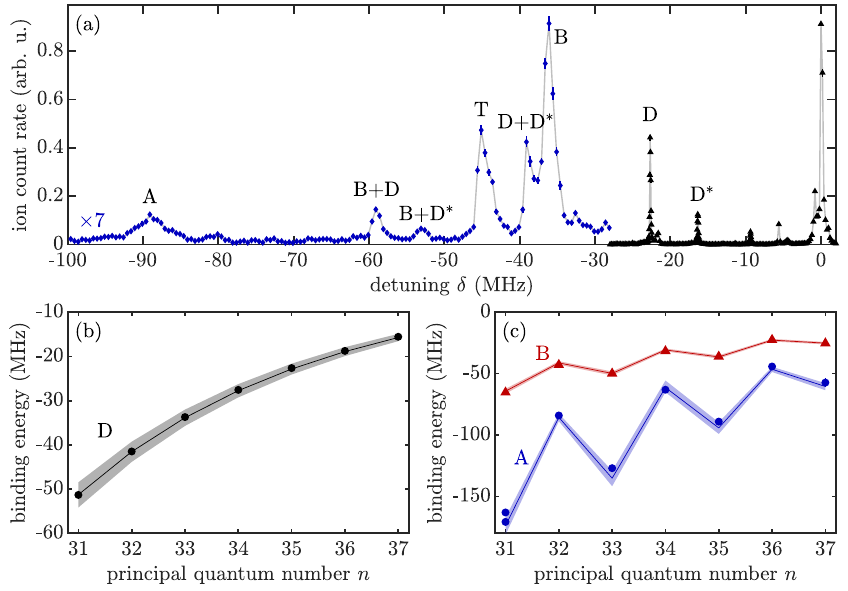}
	\caption{ULRM spectroscopy for extracting $s$- and $p$-wave scattering lengths. (a) Ion signal as a function of detuning $\delta$ from the atomic Rydberg line $|35S_{1/2},m_j=1/2\rangle$. Deeply bound molecular states are magnified for better visibility. Solid lines connect the datapoints to guide the eye. (b) Binding energy of the outermost dimer D as a function of $n$. (c) Binding energies of the deeply bound dimers A (circles) and B (triangles) as a function of $n$. Solid lines show results from a Green's function calculation with fitted $s$- and $p$-wave scattering lengths (see text). The shaded areas mark small variations on the scattering lengths as described in the text. Error bars in (a) and all other spectra denote one standard deviation. Error bars for the measured binding energies in (b) and (c) are smaller than the data points.}
	\label{Fig2}
\end{figure}

According to the above considerations, the potential detection of spin-orbit interaction in the $(^3 P_J)$ negative-ion system requires investigation of deeply-bound dimer states with sufficient resonant $p$-wave scattering character. In a first set of experiments, we aim to identify and study these candidates via extensive molecular spectroscopy. To this end, we perform Rydberg spectroscopy incorporating field ionization and subsequent ion detection starting from an ultracold ($1.5 \, \mu$K) ensemble of typically $4.5 \times 10^6$ $^{87}$Rb atoms prepared in the fully spin-stretched $|F=2,m_F=2\rangle$ hyperfine state and held in a magnetic quadrupole trap. In the trap, the atoms experience a homogeneous magnetic offset field set to $B=2.2 \, \rm{G}$. We address $nS_{1/2}$ Rydberg states via two-photon spectroscopy involving the intermediate $6P_{3/2}$ state at typical intermediate detunings between $+80$ MHz and $+400$ MHz, and laser polarizations set to address the Zeeman sublevel $m_j=1/2$.

An exemplary spectrum of the ULRMs below the $|35S_{1/2},m_j=1/2;F=2,m_F=2\rangle$ asymptote is shown in Fig.~\ref{Fig2}(a). The strongest molecular line at $-22.6$ MHz corresponds to the $s$-wave dominated dimer (D) \cite{Bendkowsky2009}. Additional lines with smaller binding energy are excited dimers bound by quantum reflection \cite{SM}, previously studied in \cite{Bendkowsky2010}. For larger binding energies, we observe two comparatively broad and so far unexplored resonances labeled A and B, which we attribute to the two deeply bound dimer states depicted in Fig.~\ref{Fig1} for $n=31$. Four remaining resonances are attributed to trimer states with binding energies that match the sum of dimer lines \cite{Bendkowsky2010}. Specifically, these comprise the simplest trimer formed by two atoms in the $s$-wave dimer state (T), a trimer formed by one atom in the $s$-wave dimer and one in the dimer state B (B+D), as well as trimers formed when one perturber resides in the strongest excited dimer state (B+D$^\ast$ and D+D$^\ast$).

In order to investigate the role of spin-orbit coupling on the deeply bound dimer A, we have taken spectra as shown in Fig.~\ref{Fig2}(a) for a range of principal quantum numbers. The measured binding energies for the states D, A and B are depicted in Figs.~\ref{Fig2}(b) and (c). For the $s$-wave dominated dimer, we observe the well-known monotonic decrease of the binding energy with $n$ \cite{Bendkowsky2009}. The deeply bound states A and B, however, show a qualitatively different behavior characterized by a strong alternation of their energy with $n$. Note that state B appears as a single resonance for all $n$. The same holds for state A except for $n=31$. Here, we observe a doublet structure split by $\approx 8$ MHz.

In a next step, we perform numerical simulations based on a Fermi model, which allow us to extract triplet $s$- and $p$-wave scattering lengths from our data. To this end, we combine advantages from two different methods for simulating PECs, i.e. Green's function calculus and Hamiltonian diagonalization on a finite basis set. Briefly, the Green's function approach intrinsically provides converged results accounting for all Rydberg levels but lacks the possibility to include the full molecular spin structure \cite{Khuskivadze2002}. Full diagonalization allows us to include all relevant spin degrees of freedom \cite{Eiles2017,Markson2016,Anderson2014}, but exhibits uncertainties originating from the chosen size of the basis set \cite{Fey2016}. We stress that it is the combination of both methods which permits conclusions on a precise quantitative level by adapting the employed basis set as outlined in the following.

Starting point is the comparatively simple spin configuration investigated in our experiment, i.e. ULRMs associated with the $|35S_{1/2},m_j=1/2;F=2,m_F=2\rangle$ asymptote. For negligible spin-orbit interaction, these molecules are described by a single (triplet) $s$- and $p$-wave scattering channel \cite{Bendkowsky2010}. Importantly, in that case complications due to atomic hyperfine structure or Rydberg fine structure do not play a role \cite{Deiss2019,Boettcher2016}. The PEC and the associated vibrational molecular states are then obtained from Green's function calculations. First, we have computed the molecular states D, A, and B using $s$- and $p$-wave scattering length data from ab-initio calculations \cite{Bahrim2000,Bahrim2001} and found rather poor agreement with the data in Fig.~\ref{Fig2}(c), particularly for the $p$-wave dominated state A. Second, we adapted the $s$- and $p$-wave scattering lengths ($a_s^T(k)$ and $a_p^T(k)$) which enter the calculations, aiming for improved agreement between experiment and simulation results. For this, we employ a comparatively simple model potential to compute the $k$-dependent scattering lengths, consisting of a long-range polarization potential and a short-range adjustable hard wall \cite{SM,Markson2016,Parsons67}. Moreover, note that the $s$- and $p$-wave channel can be adjusted independently by exploiting that the binding energy of the $s$-wave dominated dimer (D) is essentially unaffected by the $p$-wave channel.

The molecular binding energies computed with the adjusted scattering lengths are depicted with solid lines in Figs.~\ref{Fig2}(b)-(c). We obtain a zero-energy $s$-wave scattering length $a_s^T(0) = -15.2 \, a_0$ and a value for the $p$-wave shape-resonance position $E_{\rm{r}}^{\rm{avg}} = 26.6 \, \rm{meV}$ \cite{footnoteEr}. In order to estimate uncertainties for these values, the range of binding energies obtained from slight changes of the scattering lengths are indicated by shaded regions. Those correspond to variations in $a_s^T(0)$ of $\pm 0.5 \, a_0$ and $E_{\rm{r}}^{\rm{avg}}$ of $\pm 0.2 \, \rm{meV}$. Note that the ($J$-averaged) resonance position predicted in Ref.~\cite{Bahrim2000} based on a two active-electron model to account for electron correlations is about 20\% smaller. Similar discrepancy has been found in photodetachment experiments of Cs$^-$ \cite{Scheer1998,Bahrim2000}. For $a_s^T(0)$, the obtained value lies between previous theoretical estimates ($-13 \, a_0$ \cite{Bahrim2000}, $-16.9 \, a_0$ \cite{Fabrikant1986}).

\begin{figure}[!t]
\centering
	\includegraphics[width=\columnwidth]{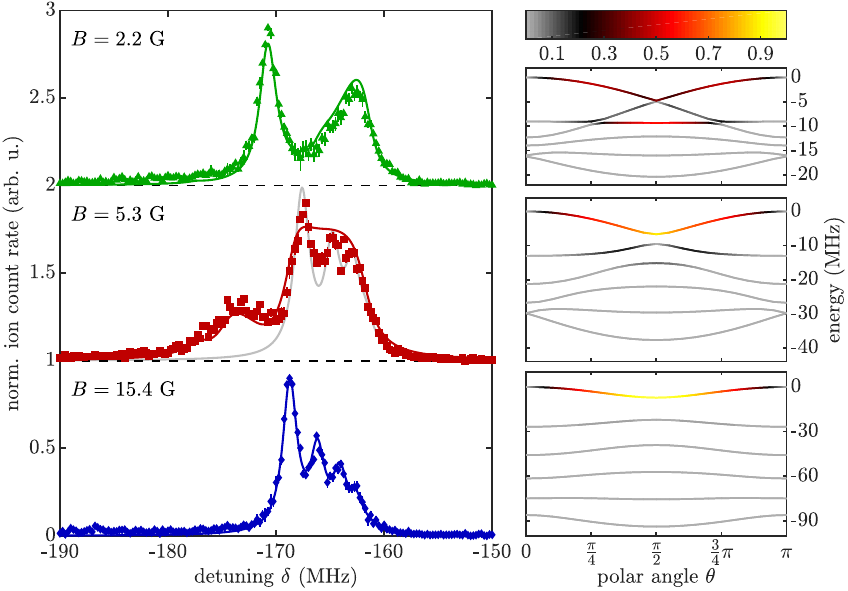}
	\caption{Spin-orbit interaction and molecular alignment. (left) Spectra of the $p$-wave dominated molecular state A with $n=31$ and for magnetic fields $B$ as indicated. Zero detuning corresponds to the atomic Rydberg line $|31S_{1/2},m_j=1/2\rangle$. Solid lines are simulated line shapes based on the $\theta$-dependent PECs \cite{SM}. (right) Angular dependence of the PECs evaluated at the minimum of the potential well in which state A is localized ($R=890 \, a_0$). Energies are referenced to the maximum of the uppermost PEC. The three plots are computed for the magnetic field present in the corresponding measurement, i.e. for increasing values of $B$ from top to bottom. Coloring encodes the projection onto $m_j=1/2$ and $m_F=2$ (see text).}
	\label{Fig3}
\end{figure}

While the Green's function calculation allows us to predict the observed molecular binding energies, it does not explain the measured doublet structure of the $p$-wave dominated state A for $n=31$. In the following, we investigate this state in more detail and demonstrate that the level splitting is directly related to the fine structure of the $^3 P_J$ negative-ion resonance. High-resolution spectroscopy of the observed doublet is shown in Fig.~\ref{Fig3} for three increasing values of the magnetic field $B$. We observe a strong qualitative change in the spectral shape when changing the magnetic field from about 2 to 15 G. While the doublet is observed for comparatively small fields (triangles), a single resonance modulated by a characteristic narrow sub-structure appears for higher values of $B$ (diamonds).

To explain this observation, we now include the full molecular spin structure into our calculation of PECs via diagonalization of the system Hamiltonian on a finite basis set \cite{Eiles2017,Markson2016,SM}. Here, the fine structure of the anion enters the computation via three $J$-dependent $p$-wave scattering channels, quantified by their corresponding scattering lengths $a_{p,J}^T (k)$. We calculate the scattering lengths using the short-range parameters obtained above from the data in Fig.~\ref{Fig2}, but now add standard $LS$-interaction to our model potential \cite{SM}, which delivers the $J$-dependent $a_{p,J}^T (k)$ and the corresponding shape-resonance positions $E_{\rm{r}}^J$. Evidently, our approach yields values for $E_{\rm{r}}^J$ which fulfill Land\'{e}'s interval rule, as expected for pure Russel-Sounders coupling \cite{Bahrim2000}. Importantly, we can largely reduce the aforementioned uncertainties arising from the choice of the basis set by switching off the anion fine structure in the calculation and then adapting the basis set to match the previous Green's function results, yielding 4 hydrogenic Rydberg manifolds with $n-5$ to $n-2$ \cite{SM}.

While the PECs shown in Fig.~\ref{Fig1} are computed for a field-free situation, the magnetic field present in the experiment renders the situation even richer. Specifically, the Zeeman energy of the electron spins lifts the pairwise degeneracy of the three $|\Omega|$-states. Furthermore, when the effect of spin-orbit coupling is sufficiently strong, the PECs obtain additional angular dependence as a result of an angular-dependent mixing of the three $p$-wave scattering channels \cite{Hummel2019}. Computed PECs for the values of $B$ set in the experiment are shown in Fig.~\ref{Fig3} (right column). For $B=2.2$ G, our laser excitation scheme couples only to the upper two PECs as indicated by the coloring, which denotes the absolute square of the projection of the electronic molecular state onto $|m_j=1/2;F=2,m_F=2\rangle$ weighted by the solid angle $\sin(\theta)$. Note that for negligible spin-orbit interaction, the PECs are independent of $\theta$ and one only couples to the highest energy state ($\Omega=+5/2$). The observed doublet is thus a direct consequence of the $^3 P_J$ fine structure.

With increasing $B$ the Zeeman shift separates the $\theta$-dependent PECs and only the curve with $\Omega=+5/2$ can be addressed. Moreover, the increasing angular confinement finally aligns the molecule, leading to a series of discrete pendular states. This transition from a doublet to a single relevant PEC which exhibits a pendular-state sub-structure (Fig.~\ref{Fig3}, $B=15.4$ G) is in excellent agreement with the experimental observation. For a quantitative comparison, the spectra are compared to simulated line shapes using a semi-classical sampling approach based on the relevant PECs in the case of unresolved pendular states ($B=2.2$ G and $B=5.3$ G) and a rigid-rotor model when individual pendular states are observed ($B=15.4$ G) \cite{SM}. Deviations for $B=5.3$ G in the spectral part associated with the uppermost PEC is due to the onset of strong molecular alignment, as seen by comparison to the rigid-rotor model prediction based on that PEC (gray line). Moreover, the excellent agreement between theory and experiment allows for extracting the fine-structure splitting of the $^3 P_J$ anion state \cite{SM}. For the fitted results shown in Fig.~\ref{Fig3}, we obtain $E_{\rm{r}}^{J=(0,1,2)} = (24.4,25.5,27.7) \, \rm{meV}$, respectively. Apart from the systematically larger value of the measured $E_{\rm{r}}^{\rm{avg}}$ discussed above, the obtained fine-structure splitting is in good agreement with the predictions in Ref.~\cite{Bahrim2000}.

\begin{figure}[!t]
\centering
	\includegraphics[width=\columnwidth]{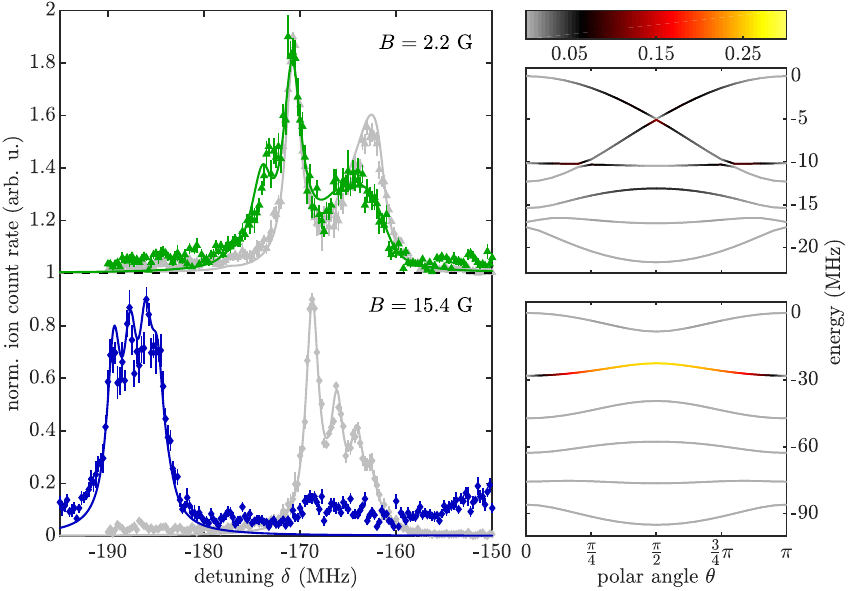}
	\caption{Spin character of the spin-orbit affected ULRM. (left) Spectra of the $p$-wave dominated molecular state A for $n=31$ with laser polarization set to address the $m_j=-1/2$ atomic Rydberg state and for magnetic fields $B$ as indicated. Gray data sets show the corresponding spectra for $m_j=1/2$ reprinted from Fig.~\ref{Fig3} for comparison. Zero detuning corresponds to the atomic Rydberg line $|31S_{1/2},m_j=1/2\rangle$. Solid lines are simulated line shapes based on the $\theta$-dependent PECs. (right) Angular dependence of the corresponding PECs similar to the ones shown in Fig.~\ref{Fig3}. Coloring now encodes the projection onto $m_j=-1/2$ and $m_F=2$.}
	\label{Fig4}
\end{figure}

Finally, we investigate the spin-character of the PECs in the presence of spin-orbit coupling by changing the laser polarization to couple to the $|31S_{1/2},m_j=-1/2\rangle$ Rydberg level. Measured spectra of the molecular state A are shown in Fig.~\ref{Fig4} for two different settings of the magnetic field. For the low-field data ($B=2.2$ G) the spectrum only slightly changes due to a small shift of the excitation strength to smaller energies. Note that this is again an effect of spin-orbit interaction, which strongly mixes the spin-character of the $\theta$-dependent PECs. For larger values of $B$, this spin-mixing is less pronounced and our excitation scheme mostly couples to the second highest energy state ($\Omega=+3/2$). This is reflected in the experiment data for $B=15.4$ G, showing a pronounced Zeeman shift of resolved pendular states. Again, we find excellent agreement with simulated line shapes.

In conclusion, we have exploited ULRMs for precise measurements on a quasi-bound negative-ion resonance. A careful analysis of measured binding energies allowed us to extract $s$- and $p$-wave scattering lengths and pinpoint the positions of the $p$-wave shape resonances associated with the $^3 P_J$ fine-structure triplet of Rb$^-$. We expect that the obtained scattering data will form the basis for future experiments on evermore delicate aspects of ULRMs, comprising few-body effects \cite{Eiles2016,Fey2019}, molecular dynamics, or more complex spin-couplings \cite{Deiss2019}. These prospects also call for developing Green's function calculations including all molecular spins. Moverover, our results allow for refining sophisticated predictions for low-energy electron-neutral scattering \cite{Bahrim2000,Bahrim2001,Khuskivadze2002}. The presented technique for measuring fine-details of near-threshold negative-ion resonances can further be transferred to benchmark other atomic and potentially also molecular systems featuring intriguing low-energy scattering properties \cite{Bartschat2003,Tarana2019,Hotop2003}.

We thank M. Eiles for numerous discussions and W. Li for assistance with the Green's function calculations.
We acknowledge support from Deutsche Forschungsgemeinschaft [Projects No. PF 381/13-1, No. PF 381/17-1, and No. SCHM 885/30-1, the latter two being part of the SPP 1929 (GiRyd)].
F. M. acknowledges support from the Carl-Zeiss foundation and is indebted to the Baden-W\"urttemberg-Stiftung for the financial support by the Eliteprogramm for Postdocs.

\clearpage

\section{Supplementary Material: Precision spectroscopy of negative-ion resonances in Rydberg molecules}

\subsection{Excited dimers bound by internal quantum reflection}

\begin{figure}[!t]
\centering
	\includegraphics[width=\columnwidth]{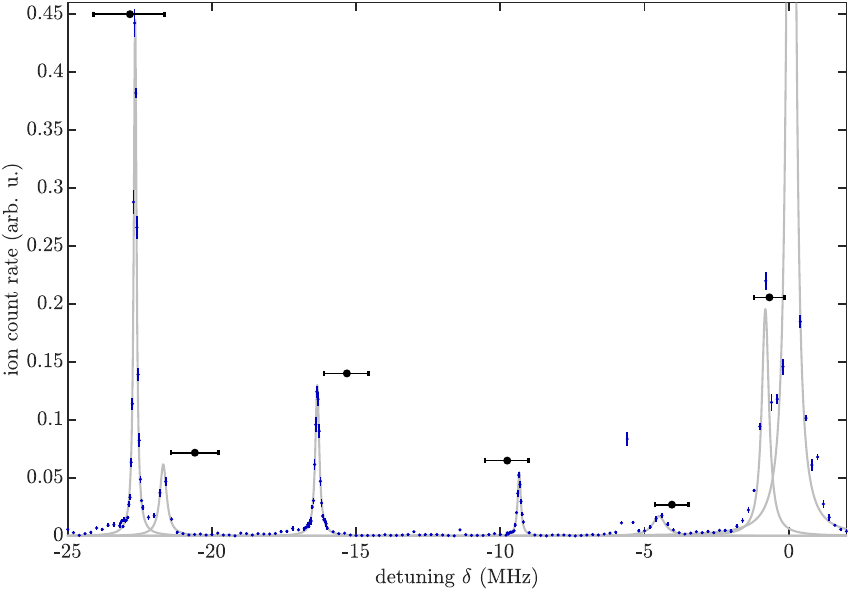}
	\caption{ULRM spectroscopy of excited dimer states bound by quantum reflection for $n=35$. Shown are the data presented in Fig.~2 of the main article with focus on the spectral region between the $s$-wave dominated dimer (D) at $\delta = -22.6$ MHz and the atomic Rydberg line $|35S_{1/2},m_j=1/2\rangle$ at $\delta=0$. The solid line is a fit to the data using a sum of multiple Lorentzians. The signal at $\delta = -5.6$ MHz stems from weak residual coupling to the Zeeman-shifted $|35S_{1/2},m_j=-1/2\rangle$ atomic Rydberg line and is excluded from the fit. Black circles are predicted molecular states from Green's function calculations with our fitted $s$- and $p$-wave scattering lengths, i.e. not accounting for spin-orbit coupling. Error bars are derived from the same variations in the scattering lengths used to obtain the shaded regions in Fig.~2.}
	\label{FigSupMat1}
\end{figure}

In the main article, we have demonstrated that the measured binding energies of the molecular states D, A, and B allow for precise fitting of the $k$-dependent (triplet) $s$- and $p$-wave scattering lengths. Here, we discuss the vibrationally excited dimer states which are formed by internal quantum reflection at the steep drop of the PEC caused by the $p$-wave shape resonance \cite{Bendkowsky2010MAT}. These states lie energetically above the vibrational ground-state dimer D. Fig.~\ref{FigSupMat1} depicts the relevant part of the spectrum presented in Fig.~2 of the main article. In total, we identify five excited dimer states lying energetically between state D at $\delta=-22.6$ MHz and the atomic Rydberg line ($\delta=0$). Note that three of the resonances ($\delta=-21.7$ MHz, $\delta=-4.5$ MHz, and $\delta=-0.8$ MHz) were not resolved in earlier work \cite{Bendkowsky2010MAT}. All the observed states are predicted by our Green's function calculations (black circles) using the adapted $a_s^T(k)$ and $a_p^T(k)$ obtained from the fitting procedure to the data of Fig.~2 (b) and (c). The good quantitative agreement provides additional support for the obtained scattering lengths. Small residual deviations could be explained by the neglected spin-orbit coupling in the Green's function calculation. 

\subsection{Model for computing electron-neutral scattering lengths}

For the calculation of the energy-dependent (triplet) $s$- and $p$-wave electron-neutral scattering lengths, we employ a comparatively simple model potential describing the Rb-e$^-$ interaction (in atomic units)
\begin{equation}
V_L (r) = - \alpha/(2 r^4) +  L(L+1)/(2 \mu_e r^2) \, .
\label{Eq_escatt}
\end{equation}
Here, $\alpha$ denotes the Rb ground-state polarizability, $r$ is the distance of the electron from the Rb core, $\mu_e$ the reduced mass, and the angular momentum $L=0$ ($L=1$) for the $s$-wave ($p$-wave) scattering channel. The short-range details are captured by an inner hard-wall at an adaptable distance $r_0$. Solving the radial Schr\"{o}dinger equation for a range of momenta $k$ yields $s$- and $p$-wave (triplet) scattering phase shifts $\delta_s^T(k)$ and $\delta_p^T(k)$, respectively. The scattering lengths are then given by $a_s^T(k) = -\tan(\delta_s^T(k))/k$ and $a_p^T(k) = -\tan(\delta_p^T(k))/k^3$. Spin-orbit interaction in the $p$-wave scattering channel is modeled by adding standard $LS$-coupling 
\begin{equation}
V_{L,S} (r) = - \beta  \frac{dV_0 (r) / dr}{2 c^2 r} \vec{L} \cdot \vec{S} \, .
\label{Eq_LS}
\end{equation}
For triplet scattering, the total electron spin $S=1$. Consequently, Eq.~\ref{Eq_LS} gives rise to three $p$-wave scattering channels with total angular momentum $\vec{J}=\vec{L}+\vec{S}$ ($J \in \{0,1,2\}$). Including Eq.~\ref{Eq_LS} into the computation of scattering phase shifts and scattering lengths yields $J$-dependent results $\delta_{p,J}^T(k)$ and $a_{p,J}^T(k)$. Each channel exhibits a shape-resonance associated with the corresponding Rb$^-$$(^3 P_J)$ negative-ion resonance. The resonance position $E_{\rm{r}}^J$ is defined as the inflection point of $a_{p,J}^T(k)$. We use the parameter $\beta \approx 1.0$ in our fitting procedure to fine tune the strength of the spin-orbit coupling for matching the observed line shapes in Fig.~3 and Fig.~4 of the main article. 

Before adjusting the short-range hard wall position $r_0$ to fit the scattering lengths to the experimental data, we have verified that our model potential reproduces the full $k$-dependence of predicted $s$- and $p$-wave phase shifts \cite{Bahrim2000MAT,Khuskivadze2002MAT}. Moreover, the functional $k$-dependence is found insensitive to the precise value of $\alpha$, i.e. for a small variation of $\alpha$ one finds a slightly shifted $r_0$ which produces the same phase shifts.

The phase shifts obtained in this work by fitting to the experimental data (see below) are shown in Fig.~\ref{FigSupMat2}, and are compared to previous theoretical predictions.

\subsection{Fitting the $k$-dependent scattering lengths to the data}

Our procedure to fit the scattering lengths $a_s^T(k)$ and $a_p^T(k)$ to the measured binding energies shown in Fig.~2 of the main article exploits the different sensitivity of the investigated molecular states to the $s$- and $p$-wave scattering channel. Starting with the $s$-wave dominated dimer (D) allows for adjusting $a_s^T(k)$ largely independent of $a_p^T(k)$. The $k$-dependent scattering length is adjusted by small variations of the hard inner wall $r_0$ aiming for minimizing the deviation between experiment and theory. Having fixed $a_s^T(k)$, we continue adjusting $a_p^T(k)$ in the same way, now minimizing deviations between measurement and calculated binding energies for the states A and B. For this, we consider all measured binding energies for state B but select a subset for state A, specifically $n=32$, $34$, and $36$. Those are the principal quantum numbers for which the state A is less bound (\textit{cf.} Fig.~2(c)), and essentially unaffected by spin-orbit coupling. This allows us to model the data with the Green's function calculation. The shaded regions in Fig.~2(c) are obtained by small variations of $r_0$, which result in the bounds on $a_s^T(0)$ and $E_{\rm{r}}^{\rm{avg}}$ given in the main article.

\begin{figure}[!t]
\centering
	\includegraphics[width=\columnwidth]{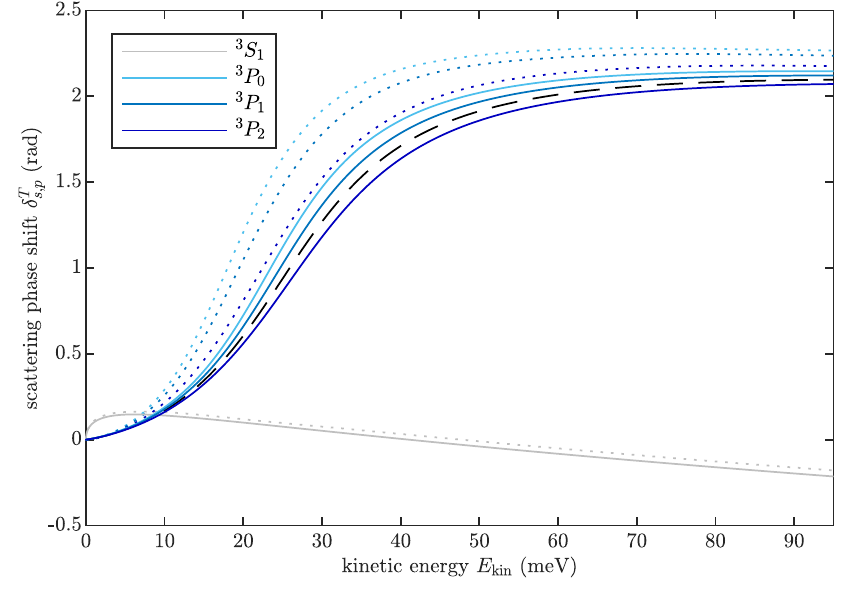}
	\caption{Rb-e$^-$ scattering phase shifts for triplet $s$- and $p$-wave scattering. The solid lines show the data for $\delta_{s}^T$ and $\delta_{p,J}^T$ as a function of collision energy $E_{\rm{kin}} = k^2/2$ obtained in this work by fitting to the measured molecular states. The dashed line denotes the $p$-wave phase shift in the absence of spin-orbit coupling used in the Green's function calculations. For comparison, the dotted lines show the predicted phase shifts reported in Refs.~\cite{Bahrim2000MAT,Khuskivadze2002MAT}.}
	\label{FigSupMat2}
\end{figure}

\subsection{Calculation of molecular potential energy curves}

\subsubsection{Hamiltonian diagonalization approach}

In the spirit of the Born-Oppenheimer approximation, we assume the nuclear motion of the molecule's constituents to be separable from the electronic motion of the Rydberg atom. Solving the stationary Schr\"odinger equation for the electronic degree's of freedom provides the potential energy curves (PECs) which are used as input for the vibrational Schr\"odinger equation. To calculate the PECs, we employ the electronic Hamiltonian (in atomic units) \cite{Hummel2017MAT}
\begin{equation}
H = H_{R} + H_{G} + H_{B} + V\,.
\label{eqn:hamiltonian}
\end{equation}
$H_{R}$ describes the dynamics of the Rydberg electron at position $\vec{r}$ in the potential of the ionic core of the Rydberg atom, which is located at the coordinate origin. The electron has spin $\vec{s}_1$ and angular momentum $\vec{l}$. Eigenstates of $H_{R}$ are $\phi_{nljm_j}(\vec{r})$ with eigenvalues $E_{nlj}$, $n$ being the principal quantum number and $j=|\vec{l}+\vec{s}_1|$ the total angular momentum of the Rydberg electron. The energies $E_{nlj}$ are taken from spectroscopic measurements \cite{Li2003MAT, Han2006MAT, Sansonetti2006MAT} and are used as input to analytically determine the long-range behavior of $\phi_{nljm_j}(\vec{r})$ in terms of appropriately phase shifted Coulomb wave functions. $H_{G}=A\,\vec{I}\cdot\vec{s}_2$ represents the Hamiltonian of hyperfine interaction in the ground-state atom with the spin of the valence electron $\vec{s}_2$, the nuclear spin $\vec{I}$,  and $A=3.417$ GHz \cite{Arimondo1977MAT}. Eigenstates of $H_{G}$ are $\ket{F m_F}$, where $F=|\vec{I}+\vec{s}_2|$. $H_{B} = \vec{B}\cdot (\vec{s}_1 + \vec{s}_2 + \vec{l}/2)$ models the Zeeman coupling of the electronic angular momenta to the magnetic field, where $\vec{B}$ is given in units of $2.35\times10^9$ G. $V$ describes the interaction between the Rydberg electron and the ground-state atom, which depends on the total electronic spin $\vec{S}=\vec{s}_1+\vec{s}_2$ as well as the orbital angular momentum $\vec{L}$ of the Rydberg electron in the reference frame of the ground-state atom. We consider singlet ($S=0$) and triplet ($S=1$) interaction for the $s$-wave ($L=0$) and $p$-wave ($L=1$) channel. To this end, we employ a generalized Fermi pseudopotential \cite{Eiles2017MAT}
\begin{equation}
V= \sum_{\beta} \frac{(2L+1)^2}{2} a_{(s ; p,J)}^{(T ; S)}(k) \frac{\delta(X)}{X^{2(L+1)}} \ket{\beta} \bra{\beta}\, .
\label{eqn:pseudopotential}
\end{equation}
Here, $X=|\vec{r}-\vec{R}|$ is the distance between the Rydberg electron and the ground-state atom and $\beta$ is a multi index that defines projectors onto the different interaction channels $\ket{\beta}=\ket{(LS)JM_J}$, with $\vec{J}=\vec{L}+\vec{S}$ being the total angular momentum of the two electrons with respect to the ground-state atom's core and $M_J$ the corresponding magnetic quantum number. The scattering lengths $a_{(s ; p,J)}^{(T ; S)}(k)$ are derived from the respective phase shifts as discussed above. Note that the upper indices $T$ and $S$ denote triplet and singlet scattering and the latter is not to be confused with the total electron spin. The wave number is calculated via the semi-classical relation $k=\sqrt{2/R-2 E^\star}$ given in terms of the energy $E^\star$ of the asymptotic atomic level $nS_{1/2}$ that we are interested in.

Note that $V$ neither commutes with $H_{R}$, $H_{G}$, nor with $H_{B}$, however, in the absence of a magnetic field, a good quantum number to discriminate the PECs is  $\Omega=m_l+m_1+m_2+m_I$, which corresponds to the projection of the total angular momentum of the (non-rotating) molecular system onto the internuclear axis. Here, $m_l, m_1, m_2, m_I$ are the magnetic quantum numbers of $l, s_1, s_2, I$, respectively. 

Considering the symmetries of a dimer in a magnetic field leads to two relevant spatial degrees of freedom: The internuclear distance $R$, and the relative angle between the magnetic field axis and the internuclear axis $\theta$. Without loss of generality, we fix the internuclear axis to be the $z$-axis such that $\vec{R}=R\hat{e}_z$ and consider rotations of the magnetic field vector around the $y$-axis such that $\vec{B}=B(\cos\theta\,\hat{e}_z + \sin\theta\,\hat{e}_x)$.

We obtain PECs by diagonalizing $H$ in a finite basis set. As stated in the main article, quantitative uncertainties in the PECs arising from the choice of the basis set are largely reduced by switching off the fine structure in the triplet scattering channel and then comparing the results to the Green's function calculations. Specifically, we find optimal matching of the two methods using a basis set which comprises in total four manifolds of electronic Rydberg states with principal quantum numbers such that two hydrogenic manifolds lie energetically below and two above the $nS_{1/2}$ state we are interested in. Further, all total angular momenta $j$ are considered, while the projections $m_j$ are truncated to neglect $|m_j|>3/2$ which do not interact with the ground-state atom. The basis $\ket{F m_F}$ is considered completely.

Additionally to the eigenvalues of $H$ constituting the PECs, by aid of the associated eigenvectors $\ket{\Psi_\epsilon(R,\theta)}$, we obtain the squared electronic dipole transition elements $|d(R,\theta)|^2 = \braket{\Psi_\epsilon(R,\theta)|\hat{P}|\Psi_\epsilon(R,\theta)}$, with $\hat{P}=| m_j;F,m_F \rangle \langle m_j;F,m_F |$, which are used for the simulation of the measured spectral line shapes (see below).

\subsubsection{Green's function approach}

Due to the mentioned convergence issues, which are inherent to the above presented diagonalization scheme in the truncated Hilbert space \cite{Fey2016MAT,Eiles2017MAT}, we also derive PECs employing alternative Green's function methods \cite{Khuskivadze2002MAT, Bendkowsky2010MAT}. These Green's function methods make use of the analytically known Coulomb Green's function $G_c(\vec{r},\vec{r}',E)$ \cite{Hostler1963MAT}, which satisfies $(-\Delta/2  - 1/r -E)G_c(\vec{r},\vec{r'},E)=\delta(\vec{r}-\vec{r'})$, where $E$ is the energy. Based on this Coulomb Green's function, one can construct a Green's function $G(\vec{r},\vec{r'},E)$ for the Rydberg electron that incorporates quantum defects $\Delta_l$ characteristic for the Rydberg atom as $G(\vec{r},\vec{r'},E)=G_c(\vec{r},\vec{r'},E)+G_{qd}(\vec{r},\vec{r'},E)$ \cite{Davydkin1971MAT}. 
 
Green's function approaches are more accurate than the truncated diagonalization in the sense that, firstly, the Green's function contains information on all bound and continuum states of the system and, secondly, it allows for a proper handling of the singular $\delta$-interaction in the pseudopotential. On the other hand Green's function methods typically neglect certain spin-interaction effects such as the fine structure of the Rydberg atom and the hyperfine structure of the ground-state atom and their coupling by the electron scattering, which is in many cases essential for a correct interpretation of spectroscopic results but has so far only been included in the framework of truncated-diagonalization schemes \cite{Eiles2017MAT}.
   
However, for the Rydberg $S$-state investigated in the experiment, the Rydberg fine structure is only of minor importance. Furthermore, the molecular states investigated in this work are essentially pure $F=2$ and pure triplet ($S=1$) states and are not affected by mixing between $S=0$ and $S=1$ states. For this reason a Green's function approach that neglects the fine structure of the Rydberg atom and takes only triplet scattering into account is well suited to describe the investigated molecular states. 

\begin{widetext}
An appropriate pseudopotential for the electron-atom interaction is in this case given by \cite{Omont1977MAT,Hamilton2002MAT}
\begin{equation}
V= 2 \pi a_s^T(k) \delta(\vec{R}-\vec{r}) + 6 \pi a_p^T(k) \overleftarrow{\nabla}_{\vec{r}} \cdot \delta(\vec{R}-\vec{r})  \overrightarrow{\nabla}_{\vec{r}} \, .
\label{eqn:pseudo_simple}
\end{equation}
The Coulomb Green's function can be written as \cite{Hostler1963MAT}
\begin{equation}
G_C(\vec{r},\vec{r'},E)= \frac{\Gamma\left(1-n^* \right)}{ \pi n^* (\xi-\eta)} \left[M'_{n^*,1/2}\left(\eta \right) W_{n^*,1/2}\left( \xi \right) 
-M_{n^*,1/2}\left(\eta \right) W'_{n^*,1/2}\left(\xi \right) \right]
\end{equation}
with $E=-1/(2n^*)$, $\xi= (r+r'+|\vec{r}-\vec{r'}|)/n^*$, $\eta= (r+r'-|\vec{r}-\vec{r'}|)/n^*$ and the Whittaker functions $W_{n^*,l+1/2}(\xi) $ and $M_{n^*,l+1/2}(x)$.
The correction term that incorporates the quantum defects is given by \cite{Davydkin1971MAT}  
\begin{equation}
G_\text{qd}(\vec{r},\vec{r'},E) = \sum \limits_{lm}
 \frac{\Gamma\left(l+1-n^*\right)\sin\pi \left(\Delta_l+l\right)}{\Gamma\left(l+1+n^*\right)\sin\pi\left(\Delta_l+n^*\right)} \frac{n^*}{rr'} W_{n^*,l+\frac{1}{2}}\left(\frac{2r}{n^*}\right) W_{n^*,l+\frac{1}{2}}\left(\frac{2r'}{n^*}\right)Y_{lm}(\hat{r})Y^*_{lm}(\hat{r'}) \, .
\end{equation}
We use quantum defects $\Delta_0=3.1314$, $\Delta_1=2.65$, $\Delta_2=1.35$, $\Delta_3=0.02$ and $\Delta_l=0$ for $l\geq4$.

Our approach to derive an equation that determines the PECs based on the Green's function follows Refs.~\cite{Andreev1984MAT,Andreev1985MAT}.
First we express the electron wave function with energy $E$ close to the position of the ground-state atom asymptotically as
\begin{equation}
\psi(\vec{X})= \sum \limits_{L=0}^{1}\sum \limits^{L}_{M=-L} c_{LM}(E) \left[ X^{-L-1} Y_{LM}(\hat{X})+ \dots+B_L(E) \left(X^L Y_{LM}(\hat{X}) + \dots \right) \right]
\label{eqn:Psi_smallX}
\end{equation}
where $\vec{X}=\vec{r}-\vec{R}$, $X$ is small, $c_{LM}(E)$ are energy-dependent coefficients and $B_L(E)= [(2L+1)a_L(E)]$ is linked to the energy-dependent scattering length and volume $a_0(E)=a_s^T(k)$ and $a_1(E)=a_p^T(k)$, respectively, where $k^2/2 - 1/R = E$. To determine the energy-dependent coefficients $c_{LM}(E)$ the above wave function needs to be matched to a solution valid at large $X$. This solution can be expressed in terms of the Green's function. Using the Lippmann Schwinger equation, $\Psi(\vec{r})=-\int d^3r' G(\vec{r},\vec{r'},E) V \psi(\vec{r'})$, and the expression for the potential $V$, Eq.~(\ref{eqn:pseudo_simple}), we obtain
\begin{equation}
\Psi(\vec{r}) = -2 \pi a_s^T(k) G(\vec{r},\vec{R},E) \Psi(\vec{R})   - 6 \pi  a_p^T(k)
\overrightarrow{\nabla}_{\vec{R}}G(\vec{r},\vec{R},E) \cdot  \overrightarrow{\nabla}_{\vec{R}}\Psi(\vec{R}) \, .
\label{eqn:lippschwing}
\end{equation}
The expressions $\Psi(\vec{R})$ and $\overrightarrow{\nabla}_{\vec{R}}\Psi(\vec{R})$ can be viewed as energy-dependent coefficients that need to be determined by matching (\ref{eqn:lippschwing}) to (\ref{eqn:Psi_smallX}) in the limit $X\to 0$.
An alternative approach to derive an equation similiar to (\ref{eqn:lippschwing}) that does not make use of the pseudopotential is provided in \cite{Andreev1985MAT}.

To simplify the notation we assume, without loss of generality, that $\vec{R}=R \vec{e}_z$. Due to the cylindrical symmetry, the magnetic quantum number $M=m$ is in this case conserved. Since we are interested in Rydberg $S$-states we focus on the symmetry subspace $M=m=0$. In that case, all cartesian components of $\overrightarrow{\nabla}_{\vec{R}}\Psi(\vec{R})$ except for the $z$-component vanish. Hence, equation (\ref{eqn:lippschwing}) becomes
\begin{equation}
\psi(\vec{X})= 2 \pi \tilde{c}_{00}(E)  G(\vec{r},\vec{R},E) + 2 \pi\tilde{c}_{10}(E)  \left. \frac{\partial}{\partial z'}G(\vec{r},\vec{r'},E)\right|_{\vec{r'}=\vec{R}} \, ,
\label{eqn:lippschwing_simplified}
\end{equation}
where $\tilde{c}_{00}(E)$ and $\tilde{c}_{10}(E)$ are energy-dependent coefficients that replace corresponding prefactors in (\ref{eqn:lippschwing}).

To match (\ref{eqn:lippschwing_simplified}) to (\ref{eqn:Psi_smallX}) for small $X$, we need to expand the expression of the Green's function in (\ref{eqn:lippschwing_simplified}) for small $X$.
This leads to \cite{Andreev1985MAT}
\begin{equation}
2 \pi G(\vec{r},\vec{R},E)=X^{-1} Y_{00}(\hat{X})+ \dots+ \sum \limits_{L'=0}^{1} A_{0L'}(E) \left(X^{L'} Y_{L'0} (\hat{X}) + \dots \right)
\end{equation}
and
\begin{equation}
2 \pi \left. \frac{\partial}{\partial z'}G(\vec{r},\vec{r'},E)\right|_{\vec{r'}=\vec{R}} =X^{-2} Y_{10}(\hat{X})+ \dots+ \sum \limits_{L'=0}^{1} A_{1L'}(E) \left(X^{L'} Y_{L'0} (\hat{X}) + \dots \right)
\end{equation}
with four energy-dependent coefficients $A_{LL'}(E)$ that satisfy $A_{LL'}(E)=A_{L'L}(E)$. 
Matching the irregular parts (divergent for small $X$) of (\ref{eqn:lippschwing_simplified}) and (\ref{eqn:Psi_smallX}) immediately yields $\tilde{c}_{LM}(E)=c_{LM}(E)$, while matching the regular parts yields the system of equations $B_L c_{L0}= \sum \limits_{L'=0}^{1} A_{L'L}(E)c_{L'0} $. 
Non trivial solutions exist only if the determinant of this system vanishes. This implies 
\begin{equation}
-A_{10}(E)^2 +  A_{11}(E) + \left(1/a_s^T(k) +A_{00}(E)\right)  \left(1/(3a_p^T(k)) +A_{11}(E)\right)=0 \, .
\label{eqn:determinant}
\end{equation}

This equation is the central result of the Green's function approach. The coefficients $A_{LL'}$ depend not only on the energy $E$ but implicitly also on the position of the ground-state atom $\vec{R}$. Using a numerical root-finding algorithm that provides solutions $E$ of (\ref{eqn:determinant}) as a function of $\vec{R}$ yields the PECs.  Knowledge of the coefficients $A_{LL'}$ is crucial for this purpose. We find
\begin{equation}
A_{00}= 2 \pi \frac{\partial}{\partial X} \left. X G\left(\vec{R}+X \vec{e}_z,\vec{R},E \right) \right|_{X=0}\, ,
\end{equation}
\begin{equation}
A_{10}= 2 \pi \frac{\partial}{\partial (X \cos \theta)} \left. G\left(\vec{R}+X(\cos \theta \vec{e}_z+ \sin \theta \vec{e}_x), \vec{R},E \right) \right|_{\theta=0,X=0} \, ,
\end{equation}
and
\begin{equation}
A_{11}= \frac{\pi}{3} \frac{\partial^3}{\partial X^3} \left. X^2 \frac{\partial}{\partial z'} G\left(\vec{R}+X \vec{e}_z,z' \vec{e}_z,E \right) \right|_{z'=R,X=0}\, .
\end{equation}
\end{widetext}

\subsection{Modeling of the spectral line shapes \label{scn:lineshape}}

In order to simulate the spectral line shape of the molecular state A (\textit{cf.} Figs.~3 and 4), we use the angular-dependent PECs $V(\theta)$ shown in Figs.~3 and 4 of the main text, which are obtained by fixing the radial coordinate $R$ to the position of the minimum $R_0$ of the potential well in which state A localizes. Then, we employ the rotational Hamiltonian
\begin{equation}
 H_{r}=\frac{\hat{N}^2}{2\mu R_0^2} + V(\theta)   
\end{equation}
with the rotational angular momentum operator $\hat{N}$ and the reduced mass $\mu$ of the diatomic system. This is the Hamiltonian of a rigid rotor. This approach is justified due to the fact that in our case, the energy scale of radial excitation is much larger then the energy scale of angular excitation. $H_{r}$ has eigenstates $\chi_\nu(\theta)$ with eigenvalues $E_\nu$ which can be obtained by diagonalizing $H_{r}$ in a basis of Legendre polynomials such that $\chi(\theta)= \sum_N c_N P_N(\cos\theta)$ or alternatively by employing standard methods such as finite difference or discrete variable representation.

Each eigenstate contributes to the observed line shape according to the Franck-Condon overlap $\Gamma \propto |\int d\theta \sin\theta \chi_\nu(\theta) d(R_0,\theta) \chi_\text{in}(\theta) |^2$, where the initial state $\chi_\text{in}(\theta)$ is assumed to be isotropic, i.e. independent of $\theta$, and $d(R_0,\theta)$ is the electronic dipole transition element derived from the solution of the electronic Hamiltonian $H$ (see above). Note that only states with even $\nu$ contribute. To compare the solution to experimental spectra, we use three fit parameters. First, we convolute eigenstates $\chi_\nu(\theta)$ with a Lorentzian line shape of constant width, which reflects the lifetime of the radial quantum reflection state. Second, we allow for an overall frequency offset (blue-shift) accounting for the radial zero-point energy. Third, the amplitude of the obtained line shape is rescaled to match the experimental results.

The rigid rotor model provides an excellent description for a sufficiently large magnetic field, separating the angular PECs $V(\theta)$ which are coupled by spin-orbit interaction. In that case the Born-Oppenheimer approximation holds and non-adiabatic electronic couplings are suppressed. This is the case for the largest applied magnetic field $B=15.4$ G. For intermediate field strength, i.e.\, $B=5.3$ G, we find that the rigid-rotor model provides only partial agreement. Specifically, the spectral part associated with the uppermost PEC is still modeled well (\textit{cf.} Fig.~3, gray line), however, we find quantitative deviations for the region below $-170$ MHz.  

\newpage

Therefore, we employ a sampling technique, which treats rotational degrees of freedom classically and is capable of reproducing the overall line shape for $B=5.3$ G and $B=2.2$ G. To this end, a random angle $\theta$ is drawn from a distribution representing an isotropic gas $p(\theta)=\sin\theta$ and the energy for this angle $V(\theta)$ is weighted by the squared electronic dipole element $|d(R_0,\theta)|^2$. We repeat this step ten thousand times to obtain a histogram, which then serves as input for the same procedure introduced for the rigid-rotor model employing the three fit parameters, i.e. to account for the molecule's lifetime, the zero-point energy, and the experimental signal amplitude.

\clearpage

\end{document}